\documentclass[%
 preprint, showkeys,showpacs, amsmath,amssymb,11pt
]{revtex4-1}

\usepackage{graphicx}
\usepackage{dcolumn}
\usepackage{bm}

\usepackage[
margin=1in,
]{geometry}

\usepackage{hyperref, mathrsfs,slashed}

\def\tr{\textrm{tr}}
\def\Tr{\textrm{Tr}}

\begin{document}

\title{\Large Non-Perturbative One-Loop Effective Action \\ for QED with Yukawa Couplings}

\author{\bf Theodore N. \surname{Jacobson}}
\email{tjacobs1@macalester.edu}
\email{jaco2585@umn.edu}
 \affiliation{{\footnotesize Department of Physics and Astronomy, Macalester College, Saint Paul, MN 55105, USA\\
 School of Physics and Astronomy, University of Minnesota, Minneapolis, MN 55455, USA}}
\author{\bf Tonnis \surname{ter Veldhuis}}%
 \email{terveldhuis@macalester.edu}
\affiliation{{\footnotesize
Department of Physics and Astronomy, Macalester College, Saint Paul, MN 55105, USA
}}

\date{\today} 


\begin{abstract}
We derive the one-loop effective action for scalar, pseudoscalar, and electromagnetic fields coupled to a Dirac fermion in an extension of QED with Yukawa couplings. Using the Schwinger proper-time formalism and zeta-function regularization, we calculate the full non-perturbative effective action to one loop in the constant background field approximation. Our result is non-perturbative in the external fields, and goes beyond existing results in the literature which treat only the first non-trivial order involving the pseudoscalar. The result has an even and odd part, which are related to the modulus and phase of the fermion functional determinant. The even contribution to the effective action involves the modulus of the effective Yukawa couplings and is invariant under global chiral transformations while the odd contribution is proportional to the angle between the scalar and pseudoscalar couplings. In different limits the effective action reduces either to the Euler-Heisenberg effective action or the Coleman-Weinberg potential. We also comment on the relationship between the odd part of the effective action and the chiral anomaly in QED.
\end{abstract}

\maketitle



\section{Introduction}

The effective action provides rich insight into the low energy regime of an underlying quantum field theory. The earliest example of a full non-perturbative effective action is the Euler-Heisenberg effective action, which describes the non-linear quantum corrections to the classical Maxwell theory, in the one-loop approximation \cite{Heisenberg:1935qt,Weisskopf:1996bu,doi:10.1142/S0217751X1430052X}. Schwinger's calculation of the same effective Lagrangian makes use of Fock's proper-time formalism and re-frames the pair production rate as a signal of the instability of the QED vacuum \cite{Fock,Schwinger}. The pair production rate can be read off from the imaginary part of the Euler-Heisenberg lagrangian, which vanishes to all orders in perturbation theory. Hence, the non-perturbative nature of the calculation captures essential physical phenomena, supplementing perturbative studies \cite{Dunne1,Dittrich:2000zu}.

The Euler-Heisenberg result, which is obtained assuming the background electromagnetic field is constant (i.e. slowly varying), has been extended to other solvable backgrounds \cite{PhysRevD.52.2422}, to supersymmetric QED, to second order in the loop expansion \cite{Clark:1987et,Kuzenko:2003qg,Kuzenko:2007cg,Dunne1,Dunne:2001pp}, and to QED in gravitational backgrounds \cite{Fucci:2009je,PhysRevD.22.343,Bastianelli:2008cu,PhysRevD.50.909}. For a review of the original Euler-Heisenberg effective action and its extensions, as well as its historical development, see \cite{Dunne1,Dunne:2012vv}. 

The Fock-Schwinger proper-time approach and its generalizations, which appear in both perturbative and non-perturbative calculations of effective Lagrangians, are especially useful because they are symmetry-preserving and can be applied to theories involving the totally antisymmetric tensor $\epsilon_{\mu_1\cdots\mu_n}$ \cite{McKeonSherry}. The proper-time formalism has been generalized to heat-kernel, zeta-function, and operator regularization, which provide powerful techniques for computing and regularizing the formal determinants and inverses of functional operators \cite{dewitt1965dynamical,Dunne2,Reuter:1984kw,Mckeon:1987ea,MCARTHUR1997525}. 

The standard procedure in operator regularization begins with the heat-kernel representation for a positive operator $\mathcal O$,  
\begin{equation}
\Tr\ \mathcal O^{-z} = \zeta_{\mathcal O}(z) = \frac{1}{\Gamma(z)}\int_0^\infty ds\ s^{z-1} \Tr\ e^{-s\mathcal O}. \label{eqn:propertime}
\end{equation}
At one-loop order, deriving the effective action involves calculating $\det \mathcal O$, where $\mathcal O$ is an operator which appears in the underlying theory and involves the background fields. For cases when $\mathcal O$ is not Hermitian, let alone positive, the fermion determinant must be split into its modulus and phase, which generate ``even'' and ``odd'' contributions to the effective action, respectively \cite{McKeonSherry}. In the present work, we find a much simpler way of calculating the even and odd contributions to the effective action which circumvents the need for computing the phase of the functional determinant directly. 

In the perturbative regime, proper-time techniques have been used to compute the radiatively induced effect of adding Lorentz- and CPT-violation to QED \cite{Sitenko:2001iw,Sitenko:2002fy,PhysRevD.89.045005,Borges:2016uwl}, as well as the effective action for the Yukawa model in curved spacetime \cite{Toms:2018wpy}. In the non-perturbative regime, heat-kernel methods were used to obtain the world-line path integral for fermions with general scalar, pseudoscalar, and vector couplings \cite{DHoker:1995aat,DHoker:1995uyv}. While world-line path integrals are in the same spirit, the full closed-form effective action for QED with Yukawa couplings has not yet been derived. In the present paper, we provide a simple derivation of the effective action for fermions in the one-loop and constant background field approximations. Our result is non-perturbative in the background scalar, pseudoscalar, and electromagnetic fields. The even portion of the effective action is similar to the Euler-Heisenberg effective action, except the fermion mass is modified by the Yukawa couplings. The odd portion of the effective action is proportional to the CP-odd Lorentz scalar $F\tilde F$ and the angle between the effective scalar and pseudoscalar Yukawa terms.


\section{The model} \label{sec:themodel}

We consider a simple model consisting of a Dirac fermion $\psi$ coupled to a background scalar $\Phi$, pseudoscalar $\Pi$, and $U(1)$ gauge field $A_\mu$, with Lagrangian 
\begin{equation}
\mathscr L = \mathscr L_{\text{QED}} + \mathscr L_{\text{YUK}}. \label{eqn:lagrangian}
\end{equation}
In the $(-,+,+,+)$ signature, the QED and Yukawa Lagrangians are
\begin{subequations}
\begin{eqnarray}
\mathscr{L}_{\text{QED}} &=& -\frac{1}{4}F_{\mu\nu}F^{\mu\nu}-\frac{\xi}{2}(\partial_\mu A^\mu)^2 -\bar\psi (i\gamma^\mu D_\mu +m)\psi , \label{eqn:Lqed} \\
\mathscr{L}_{\text{YUK}} &=& -\frac{1}{2} \partial_\mu \Phi\partial^\mu\Phi - \frac{1}{2} \partial_\mu \Pi\partial^\mu\Pi - V[\Phi,\Pi] -\bar\psi (\kappa\Phi+i\gamma^5\lambda\Pi)\psi, \label{eqn:Lyuk} 
\end{eqnarray}
\end{subequations}
where $F_{\mu\nu}=\partial_\mu A_\nu - \partial_\nu A_\mu$ is the $U(1)$ field strength, the covariant derivative is $D_\mu = \partial_\mu +ie A_\mu$, and $\kappa$ and $\lambda$ are real coupling constants. The scalar potential $V[\Phi,\Pi]$ also contains possible mass terms for the scalar fields, and $\xi$ is the gauge-fixing parameter. Throughout the following we will use the Dirac slash notation $\slashed D \equiv \gamma^\mu D_\mu$. With our metric signature, the gamma matrices satisfy
\begin{equation}
\left\{\gamma^\mu,\gamma^\nu \right\} = -2g^{\mu\nu}, \quad \left\{\gamma^\mu,\gamma^5\right\} = 0,
\end{equation}
and we adopt the following definition for $\gamma^5$,
\begin{equation}
\gamma^5 = \frac{i}{4!}\epsilon_{\alpha\beta\mu\nu}\gamma^\alpha\gamma^\beta\gamma^\mu\gamma^\nu = -i\gamma^0\gamma^1\gamma^2\gamma^3, 
\end{equation}
where $\epsilon^{\alpha\beta\mu\nu}$ is totally antisymmetric with $\epsilon^{0123}=+1$. The background fields are assumed to be Hermitian, $\Phi^\dagger=\Phi$, $\Pi^\dagger = \Pi$, $A_\mu^\dagger = A_\mu$, and we assume a representation for the gamma matrices such that 
\begin{equation}
(\gamma^\mu)^\dagger = \gamma^0\gamma^\mu\gamma^0, \quad (\gamma^5)^\dagger = \gamma^5. 
\end{equation}
With these Hermiticity conditions, one can easily check that the Lagrangian density \eqref{eqn:lagrangian} is real. 

In the constant field approximation, $\Phi$ is essentially treated as a VEV which simply modifies the fermion mass $m$. Furthermore, at the classical level the pseudoscalar Yukawa coupling can be removed by transforming $\psi = e^{-i\theta \gamma^5}\psi'$ and $\bar\psi = \bar\psi' e^{-i\theta \gamma^5}$, for a suitably chosen constant $\theta$. At the quantum level this is no longer the case, and as we will see, the pseudoscalar coupling gives rise to the odd part of the effective action. We will revisit the role of the parameter $\theta$ in more detail in Subsection \ref{sec:effectivelagrangian}.


\section{One-Loop Effective Potential}

The effective action is formally defined to be
\begin{align}
\Gamma =-i \ln \int \mathscr D\psi \mathscr D\bar\psi \exp\left\{i\int d^4x \ \mathscr L \right\},
\end{align}
which, for the Lagrangian \eqref{eqn:lagrangian} and in the one-loop approximation, reads 
\begin{eqnarray}
\Gamma_{\text{1-loop}} &=&-i \ln\det \left(-i\slashed D-m-\kappa\Phi-i\gamma^5\lambda\Pi\right) = -i\Tr\ln\left(-i\slashed D-m-\kappa\Phi-i\gamma^5\lambda\Pi\right), \label{eqn:effectiveaction1}
\end{eqnarray}
where $\Tr\ \mathcal O= \tr\int d^4x\langle x | \mathcal O|x\rangle$ is a trace over Dirac indices as well as spacetime. 

Due to the presence of the pseudoscalar coupling, the operator $\mathcal O$ appearing in the theory is not Hermitian, and the standard procedure for computing functional determinants does not apply. Rather than computing the fermion determinant directly, we instead compute its contribution to the pseudoscalar current, 
\begin{eqnarray}
J_\Pi  &=& \frac{1}{i}\frac{1}{\lambda}\frac{\delta\Gamma_{\text{1-loop}}}{\delta \Pi} =i \tr\langle x |\gamma^5\left(-i\slashed D-m-\kappa\Phi-i\gamma^5\lambda\Pi\right)^{-1}|x\rangle. \label{eqn:current1}
\end{eqnarray}
This approach circumvents the more delicate splitting of the fermion determinant into its modulus and phase. Now we multiply the numerator and denominator by $-i\slashed D +m+\kappa\Phi-i\gamma^5\lambda\Pi$, 
\begin{eqnarray}
J_\Pi = &i\tr\langle x |&\left(-i\slashed D+m+\kappa\Phi-i\gamma^5\lambda\Pi\right)\gamma^5  \nonumber\\
&&\times \Big[\left(-i\slashed D-m-\kappa\Phi-i\gamma^5\lambda\Pi\right)\left(-i\slashed D+m+\kappa\Phi-i\gamma^5\lambda\Pi\right)\Big]^{-1}|x\rangle \nonumber \\
= &i\tr\langle  x |&\gamma^5\left(i\slashed D+m+\kappa\Phi-i\gamma^5\lambda\Pi\right)\Big[(i\slashed D)^2-(m+\kappa\Phi)^2-(\lambda\Pi)^2\Big]^{-1} |x\rangle, \label{eqn:current2}
\end{eqnarray}
assuming $\Phi$ and $\Pi$ are constant background fields. The denominator in \eqref{eqn:current2} is positive, and hence can be represented by a proper-time integral. Taking the limit $z\to 1$ in \eqref{eqn:propertime},
\begin{eqnarray}
J_\Pi = -i\int_0^\infty & ds\ \tr \langle & x |\gamma^5\left(i\slashed D+m+\kappa\Phi-i\gamma^5\lambda\Pi\right)e^{-s\left(-(i\slashed D)^2+(m+\kappa\Phi)^2+(\lambda\Pi)^2\right)}  |x\rangle \nonumber \\
= -i\int_0^\infty & ds\ \tr \langle & x |\gamma^5\left(m+\kappa\Phi-i\gamma^5\lambda\Pi\right) e^{-s\left(-(i\slashed D)^2+(m+\kappa\Phi)^2+(\lambda\Pi)^2\right)}  |x\rangle, 
\end{eqnarray}
where we have used the fact that $\gamma^5$ times an odd number of $\gamma^\mu$ is traceless. The current splits into two parts, which we call even and odd,  
\begin{subequations}
\begin{align}
J_\Pi^\text{even} &= -\lambda\Pi \int_0^\infty ds\  e^{-s\left((m+\kappa\Phi)^2+(\lambda\Pi)^2\right)}\tr \langle x |e^{s(i\slashed D)^2}  |x\rangle,  \\
J_\Pi^\text{odd} &= -i(m+\kappa\Phi) \int_0^\infty ds\ e^{-s\left((m+\kappa\Phi)^2+(\lambda\Pi)^2\right)}\tr\langle x |\gamma^5e^{s(i\slashed D)^2}  |x\rangle.
\end{align}
\end{subequations}
The remaining step in the calculation involves evaluating the traces
\begin{equation}
\tr \langle x |e^{s(i\slashed D)^2}  |x\rangle \text{ and } \tr\langle x |\gamma^5e^{s(i\slashed D)^2}  |x\rangle. \label{eqn:matrixelement}
\end{equation}

\subsection{Coincidence limit of the fermion propagator}

There are various methods for calculating \eqref{eqn:matrixelement} available in the literature, for instance \cite{Schwinger, Schwartz:2013pla,ItzyksonZuber,Bagrov:1529647}. For completeness, we review the calculation, loosely following the procedure found in chapter 4 of \cite{ItzyksonZuber}, evaluating \eqref{eqn:matrixelement} directly rather than taking the coincident limit of the full propagator. Those familiar with the calculation can proceed to Subsection \ref{sec:effectivelagrangian}. 

We begin by splitting $\gamma^\mu\gamma^\nu$ into symmetric and antisymmetric parts, 
\begin{equation}
\gamma^\mu\gamma^\nu = -(g^{\mu\nu}+i\sigma^{\mu\nu}), \quad \sigma^{\mu\nu} = \frac{i}{2}\left[\gamma^\mu,\gamma^\nu\right].
\end{equation}
Together with $[D_\mu,D_\nu] = ieF_{\mu\nu}$, this yields the identity
\begin{equation}
-(i\slashed D)^2 = (iD)^2 + \frac{e}{2}\sigma^{\mu\nu}F_{\mu\nu}.
\end{equation}
Since $F_{\alpha\beta}$ is constant and hence commutes with $D_\mu$, the exponential factorizes,
\begin{equation}
\langle x |e^{s(i\slashed D)^2}  |x\rangle = \langle x |e^{-s\left(\frac{e}{2}\sigma^{\mu\nu}F_{\mu\nu}\right)}e^{-s(i  D)^2}  |x\rangle. \label{eqn:matrixelement3}
\end{equation}
Using the identity
\begin{eqnarray}
\sigma^{\mu\nu}\sigma^{\alpha\beta} &=& g^{\mu\alpha}g^{\nu\beta} - g^{\mu\beta}g^{\nu\alpha} - i \epsilon^{\mu\nu\alpha\beta}\gamma^5  \label{eqn:sigmaid} \\ 
 &&+ i(g^{\mu\alpha} \sigma^{\nu\beta}-g^{\nu\alpha} \sigma^{\mu\beta} + g^{\nu\beta} \sigma^{\mu\alpha}-g^{\mu\beta} \sigma^{\nu\alpha} )\nonumber,
\end{eqnarray}
one can easily show that 
\begin{equation}
\sigma^{\mu\nu}\sigma^{\alpha\beta}F_{\mu\nu}F_{\alpha\beta} = 2 (FF-i\gamma^5 F\tilde F),
\end{equation}
where $\tilde F^{\alpha\beta} = \frac{1}{2}\epsilon^{\alpha\beta\mu\nu}F_{\mu\nu}$. Defining $X \equiv FF-i\gamma^5 F\tilde F$, we have
\begin{eqnarray}
e^{-s\left(\frac{e}{2}\sigma^{\mu\nu}F_{\mu\nu}\right)} &=& 1 + \frac{s^2}{2!}\frac{e^2X}{2} + \frac{s^4}{4!}\frac{e^4X^2}{4} + \frac{s^6}{6!}\frac{e^6X^3}{8} + \cdots \nonumber \\
&&-s\frac{e}{2}\sigma^{\mu\nu}F_{\mu\nu}- \frac{s^3}{3!}\frac{e^3}{4} X \sigma^{\mu\nu}F_{\mu\nu} + \cdots \label{eqn:sigmaFexpansion}
\end{eqnarray}
We are not interested in the traceless terms in 
\begin{equation}
e^{-s\left(\frac{e}{2}\sigma^{\mu\nu}F_{\mu\nu}\right)}, \quad \gamma^5 e^{-s\left(\frac{e}{2}\sigma^{\mu\nu}F_{\mu\nu}\right)},
\end{equation}
and because $\tr(\sigma^{\mu\nu}) = \tr(\gamma^5\sigma^{\mu\nu}) = 0$, we can disregard the odd terms in the expansion \eqref{eqn:sigmaFexpansion}. For the even terms, it is useful to expand the powers of $X$ into parts proportional to the identity and parts proportional to $\gamma^5$. To demonstrate, the first four powers of $X$ are 
\begin{eqnarray}
X &=& FF-i\gamma^5F\tilde F,  \\
X^2 &=& (FF)^2-(F\tilde F)^2 - 2i\gamma^5 (FF)(F\tilde F), \nonumber \\
X^3 &=& \left( (FF)^3-3(FF)(F\tilde F)^2 \right) + i \gamma^5 \left( (F\tilde F)^3 - 3(F\tilde F)(FF)^2 \right), \nonumber\\
X^4 &=& \left( (FF)^4 + (F\tilde F)^4 - 6(FF)^2(F\tilde F)^2 \right) + i\gamma^5 \left(4(FF)(F\tilde F)^3 - 4(F\tilde F)(FF)^3 \right).\nonumber
\end{eqnarray}
In order to re-sum this series we make an explicit choice for the form of $F_{\mu\nu}$. So long as the electric and magnetic fields are not perpendicular, we can always Lorentz-transform to a frame where they are parallel. Hence, without loss of generality, we let the electric and magnetic fields point in the $z$-direction. For $\mathbf{E} = a \hat{z}$ and $\mathbf{B} = b\hat{z}$, where $a$ and $b$ are space-time constants, the Lorentz-invariant combinations become
\begin{equation}
FF = -2(a^2-b^2), \quad F\tilde F = -4ab. \label{eqn:lorentzinvariants}
\end{equation}
Inverting these equations gives $a$ and $b$ in terms of the general Lorentz invariant quantities $FF$ and $F\tilde F$,
\begin{subequations}
\label{eqn:lorentzinvariants2}
\begin{eqnarray}
a &=& \frac{1}{2}\sqrt{\sqrt{(FF)^2+(F\tilde F)^2} - FF}, \\
b &=& \frac{1}{2}\sqrt{\sqrt{(FF)^2+(F\tilde F)^2} + FF}.
\end{eqnarray}
\end{subequations}
Once expressed in terms of $a$ and $b$, the expansion \eqref{eqn:sigmaFexpansion} can be easily re-summed,
\begin{eqnarray}
\left. e^{-s\left(\frac{e}{2}\sigma^{\mu\nu}F_{\mu\nu}\right)}\right|_{\text{even}} &=& \cos(eas)\cosh(ebs) + i\gamma^5\sin(eas)\sinh(ebs). \label{eqn:sigmaFtrace}
\end{eqnarray}
The odd terms have a simple relation to the even terms but for our present purposes they are not necessary and we shall omit them. 

Continuing with our calculation of \eqref{eqn:matrixelement3}, we need to evaluate
\begin{equation}
\langle x |e^{-s(i  D)^2}|x\rangle.
\end{equation}
Analytically continuing $s\to is$, and introducing the operator $p_\mu = i\partial_\mu$, we can identify the proper-time Hamiltonian
\begin{equation}
H = (iD)^2 = (p_\mu-eA_\mu(x))^2. 
\end{equation}
The classical gauge field configuration $A_\mu$ is a function of the space-time coordinates $x^\mu$, which satisfy canonical commutation relations,
\begin{equation}
[x_\mu,p_\nu] = ig_{\mu\nu}.
\end{equation}
Up to a gauge transformation, the nonzero components of the electromagnetic potential are $A_1 = -bx_2, A_3=-ax_0$. The Hamiltonian is thus
\begin{eqnarray}
 H &=& -p_0^2+p_1^2+p_2^2+p_3^2+e^2a^2x_0^2 +e^2b^2x_2^2 +2eb p_1x_2 + 2eap_3x_0  \nonumber \\
&\equiv&  H_{03} +  H_{12}.
\end{eqnarray}
We now use the Baker-Campbell-Hausdorff formula,
\begin{equation}
e^{A}B e^{-A} = B + [A,B] + \frac{1}{2!}[A,[A,B]] + \cdots
\end{equation}
where $A_{03} = -ip_0p_3/ea$ and $B_{03} = -p_0^2 + e^2a^2x_0^2$. Using $[p_\mu,x_\nu x^\nu] = -2ix_\mu$, we have the following, 
\begin{eqnarray}
[A_{03},B_{03}] &=& 2eap_3x_0, \\
\left[A_{03},[A_{03},B_{03}]\right] &=& p_3^2, \nonumber \\
\left[A_{03},\left[A_{03},[A_{03},B_{03}]\right]\right] &=& 0,\nonumber
\end{eqnarray}
where all successive terms in the Hausdorff expansion are zero. Thus, 
\begin{equation}
H_{03} =-e^{-ip_0p_3/ea} (p_0^2 -e^2a^2x_0^2) e^{ip_0p_3/ea}.
\end{equation}
Similarly, we find
\begin{equation}
H_{12} = -e^{ip_1p_2/eb} (p_2^2 +e^2b^2x_2^2) e^{-ip_1p_2/eb}.
\end{equation}
The original matrix element then factorizes as follows,
\begin{equation}
\langle x|e^{-is  H}|x\rangle =U_{03}U_{12},
\end{equation}
where we have defined $U_{ij} \equiv \langle x_ix_j|e^{-is H_{ij}}|x_ix_j\rangle$. We begin with $U_{03}$, inserting a complete set of momentum states,
\begin{eqnarray}
U_{03} &=& \int \frac{dp_0 dp_3 dp_0' dp_3'}{(2\pi)^4} \langle x_0x_3|p_0p_3\rangle \langle p_0p_3|e^{-is H_{03}}|p_0'p_3'\rangle \langle p_0'p_3'|x_0x_3\rangle \nonumber\\
&=& \int \frac{dp_0 dp_3 dp_0' dp_3'}{(2\pi)^4} e^{ix_0(p_0-p_0')+ix_3(p_3-p_3')}\langle p_0p_3|e^{-is H_{03}}|p_0'p_3'\rangle.
\end{eqnarray}
When we exponentiate a Hamiltonian of this form, we get
\begin{equation}
e^{-is(e^A B e^{-A})} =  e^A e^{-is B} e^{-A}.
\end{equation}
Making use of this identity and inserting another complete set of states, we have
\begin{eqnarray}
U_{03}  &=& \int \frac{dp_0 dp_3 dp_0' dp_3'dq_0 dq_3 dq_0' dq_3'}{(2\pi)^8} e^{ix_0(p_0-p_0') + ix_3(p_3-p_3')}\nonumber\\
&&\times \langle p_0p_3|e^{-i\hat{p}_0\hat{p}_3/ea}|q_0q_3\rangle\langle q_0q_3 | e^{is(\hat{p}_0^2-e^2a^2\hat{x}_0^2)}|q_0'q_3'\rangle \langle q_0'q_3'|e^{i\hat{p}_0\hat{p}_3/ea}|p_0'p_3'\rangle \nonumber\\
&=& \int \frac{dp_0 dp_3 dp_0' dp_3'dq_0 dq_3 dq_0' dq_3'}{(2\pi)^8} e^{ix_0(p_0-p_0') + ix_3(p_3-p_3')} \nonumber\\
&&\times e^{-iq_0q_3/ea}\langle p_0p_3|q_0q_3\rangle\langle q_0q_3 | e^{is(\hat{p}_0^2-e^2a^2\hat{x}_0^2)}|q_0'q_3'\rangle  e^{i p_0'p_3'/ea}\langle q_0'q_3'|p_0'p_3'\rangle \nonumber\\
&=& \int \frac{dp_0 dp_3 dp_0' dp_3'dq_0 dq_3 dq_0' dq_3'}{(2\pi)^4} e^{ix_0(p_0-p_0') + ix_3(p_3-p_3')}e^{i(p_0'p_3'-q_0q_3)/ea} \nonumber\\
&&\times \delta(p_0-q_0)\delta(p_3-q_3)\delta(p_0'-q_0') \delta(p_3'-q_3')\langle q_0q_3 | e^{is(\hat{p}_0^2-e^2a^2\hat{x}_0^2)}|q_0'q_3'\rangle.
\end{eqnarray}
Integrating over $q$ and $q'$, we have
\begin{eqnarray}
U_{03} &=& \int \frac{dp_0 dp_3 dp_0' dp_3'}{(2\pi)^4} e^{ix_0(p_0-p_0') + ix_3(p_3-p_3')} e^{i(p_0'p_3'-p_0p_3)/ea} \langle p_0p_3 | e^{i\tau(\hat{p}_0^2-e^2a^2\hat{x}_0^2)}|p_0'p_3'\rangle \nonumber\\
&=& \int \frac{dp_0 dp_3 dp_0' dp_3'}{(2\pi)^3} e^{ix_0(p_0-p_0') + ix_3(p_3-p_3')} e^{i(p_0'p_3'-p_0p_3)/ea}\delta(p_3-p_3')\langle p_0 | e^{i\tau(\hat{p}_0^2-e^2a^2\hat{x}_0^2)}|p_0'\rangle \nonumber\\
&=& \int \frac{dp_0 dp_3 dp_0'}{(2\pi)^3} e^{ix_0(p_0-p_0')}e^{ip_3(p_0'-p_0)/ea} \langle p_0 | e^{is(\hat{p}_0^2-e^2a^2\hat{x}_0^2)}|p_0'\rangle.
\end{eqnarray}
Integrating over $p_3$, we get
\begin{eqnarray}
U_{03} &=&\frac{ea}{4\pi^2} \int dp_0 dp_0' e^{ix_0(p_0'-p_0)}\delta(p_0'-p_0) \langle p_0 | e^{is(\hat{p}_0^2-e^2a^2\hat{x}_0^2)}|p_0'\rangle \nonumber \\
&=& \frac{ea}{4\pi^2} \int dp_0 \langle p_0 | e^{is(\hat{p}_0^2-e^2a^2\hat{x}_0^2)}|p_0 \rangle.
\end{eqnarray}
We recognize this as a harmonic oscillator with Hamiltonian
\begin{equation}
H_0 = p_0^2 + \omega_0^2x_0^2, \quad \omega_0 = iea,
\end{equation}
with energy spectrum
\begin{equation}
H_0|n\rangle = 2\omega_0\left(n + \frac{1}{2}\right) |n\rangle. 
\end{equation}
The matrix element we want to calculate is thus
\begin{eqnarray}
U_{03} &=& \frac{ea}{4\pi^2} \int  \sum_{n=0}^\infty dp_0 \langle p_0 |n\rangle\langle n| e^{is H_0}| n\rangle \langle n|p_0\rangle \nonumber\\
&=& \frac{ea}{2\pi} \sum_{n=0}^\infty \int \frac{dp_0}{(2\pi)} |\langle p_0 | n\rangle|^2\ \exp\left\{2i\tau \omega_0\left(n+\frac{1}{2}\right) \right\} = \frac{ea}{2\pi}e^{is\omega_0} \sum_{n=0}^\infty e^{(2is\omega_0)n}.
\end{eqnarray}
The sum is clearly the Taylor expansion of
\begin{equation}
\frac{1}{1-e^{2is\omega_0}} = \frac{1}{1-e^{-2eas}},
\end{equation}
so in the end we have
\begin{equation}
U_{03} = \frac{ea}{2\pi} \frac{e^{-eas}}{1-e^{-2eas}} = \frac{ea}{4\pi \sinh(eas)}.
\end{equation}
The case for $U_{12}$ is similar, and we will not repeat the derivation here. We note, however, that we are free to choose the positive or negative square root when defining the ``Landau" frequencies $\omega_0$ and $\omega_2$. By setting $\omega_2 = -eb$, the result matches the free-field propagator in the limits $a,b \to 0$. With this choice, the matrix element is
\begin{equation}
U_{12} = \frac{eb}{2\pi} \frac{e^{-iebs}}{1-e^{-2iebs}} = \frac{eb}{4\pi i \sin(ebs)}. 
\end{equation}
Finally, we return $s\to-is$ to  obtain our result,
\begin{equation}
\langle x |e^{-s(iD)^2}|x \rangle = -\frac{e^2ab}{16\pi^2i\sin(eas)\sinh(ebs)}. \label{eqn:propertimetrace}
\end{equation}

Combining \eqref{eqn:sigmaFtrace} and \eqref{eqn:propertimetrace}, we find 
\begin{subequations}
\begin{eqnarray}
\tr\langle x| e^{s(i\slashed D)^2}|x\rangle &=& -\frac{e^2ab}{4\pi^2i}\cot(eas)\coth(ebs), \\
\tr\langle x| \gamma^5 e^{s(i\slashed D)^2}|x\rangle &=& -\frac{e^2ab}{4\pi^2}.
\end{eqnarray}
\end{subequations}

\subsection{Effective lagrangian}
\label{sec:effectivelagrangian}

Putting the pieces together, the pseudoscalar current becomes 
\begin{subequations}
\label{eqn:currents}
\begin{eqnarray}
J_\Pi^\text{even} &=& \frac{1}{i}\frac{e^2ab}{4\pi^2}\lambda\Pi \int_0^\infty ds\ e^{-s\left((m+\kappa\Phi)^2+(\lambda\Pi)^2\right)}\cot(eas)\coth(ebs),  \\
J_\Pi^\text{odd} &=& -\frac{1}{i}\frac{e^2ab}{4\pi^2}(m+\kappa\Phi)\int_0^\infty ds\ e^{-s\left((m+\kappa\Phi)^2+(\lambda\Pi)^2\right)} = -\frac{1}{i}\frac{e^2ab}{4\pi^2}\frac{(m+\kappa\Phi)}{(m+\kappa\Phi)^2+(\lambda\Pi)^2}. \label{eqn:oddcurrent}
\end{eqnarray}
\end{subequations}
Integrating with respect to $\Pi$ and multiplying by $i\lambda$, we find the even and odd parts of the effective Lagrangian: $\mathscr L_{\text{eff}} = \mathscr{L}_{\text{eff}}^\text{even} + \mathscr{L}_{\text{eff}}^\text{odd}$, where
\begin{subequations}
\label{eqn:effectivelagrangian1}
\begin{align}
\mathscr{L}_{\text{eff}}^\text{even} &= \frac{e^2}{32\pi^2}F\tilde F \int_0^\infty \frac{ds}{s} e^{-s\left((m+\kappa\Phi)^2+(\lambda\Pi)^2\right)} \cot(eas)\coth(ebs), \label{eqn:effectivelagrangianeven1} \\
\mathscr{L}_{\text{eff}}^\text{odd} &=  \frac{e^2}{16\pi^2}F\tilde F\arctan\left(\frac{\lambda\Pi}{m+\kappa\Phi}\right). \label{eqn:effectivelagrangianodd1}
\end{align}
\end{subequations}
This is our main result. 

It is important to note that we have captured the full non-perturbative effects of the pseudoscalar coupling. Though the first-order term from $\mathscr{L}_{\text{eff}}^\text{odd}$ can be found in \cite{Schwinger}, our result is valid \emph{to all orders} in $\Pi$. The closed-form Lagrangian \eqref{eqn:effectivelagrangianodd1} also exhibits some interesting features. Most strikingly, only $F\tilde F$ appears, with no higher order corrections past $\mathcal O(e^2)$. Hence, any couplings involving higher odd powers of $F\tilde F$ (if they appear in the effective theory at all) must necessarily involve derivatives of $F_{\mu\nu}$ or $\Pi$. In addition, it is interesting to note that for very large arguments the arctangent is approximately constant and tends to $\pi/2$. Hence, when  (for vanishing $\Phi$) the ratio $\Pi/m \gg 1$, the term is nearly a total derivative.

We now return to the question of chiral invariance. While the even portion of the effective action clearly retains a global chiral symmetry, the odd portion is in fact related to the parameter $\theta$ introduced in Section \ref{sec:themodel}. To be more explicit, the pseudoscalar term in the QED-Yukawa Lagrangian can be removed with a chiral rotation where the transformation parameter $\theta$ satisfies
\begin{equation}
\theta = \frac{1}{2}\arctan\left(\frac{\lambda\Pi}{m+\kappa\Phi} \right), 
\end{equation}
which allows us to rewrite the odd portion of the effective action as
\begin{equation}
\mathscr{L}_{\text{eff}}^\text{odd} =  \frac{e^2}{8\pi^2}\theta F\tilde F.
\end{equation}
However, the parameter $\theta$ is spacetime-independent only if $\Phi$ and $\Pi$ are also spacetime-independent. As the effective action \eqref{eqn:effectivelagrangian1} represents the leading contribution in a derivative (momentum) expansion, we are treating $\Phi$ and $\Pi$ not as constants but as slowly-varying fields. From this point of view, we should treat $\theta = \theta(x)$ as a local parameter which fails to leave the kinetic term of \eqref{eqn:Lqed} invariant. Nonetheless, the identification of $\theta$ in the odd portion of the effective action sheds light on chiral invariance in the effective theory. The even part of the effective action depends on the ``modulus" $(m+\kappa\Phi)^2 + (\lambda\Pi)^2$, whereas the odd part depends on the angle itself between $m+\kappa\Phi$ and $\lambda\Pi$. 

A note about passing from the pseudoscalar current \eqref{eqn:currents} to the effective Lagrangian \eqref{eqn:effectivelagrangian1}. In the process of integration we have the freedom to add to the Lagrangian an arbitrary functional that does not depend on $\Pi$. However, we can just as easily vary \eqref{eqn:effectiveaction1} with respect to $\Phi$ or $m$ and obtain \eqref{eqn:effectivelagrangianeven1} and \eqref{eqn:effectivelagrangianodd1} up to a total derivative. Hence, we are free to add some function of $A_\mu$ only, which by dimensional grounds must either be a correction to the cosmological constant, the free photon term $F_{\mu\nu}F^{\mu\nu}$, or  $F_{\mu\nu}\tilde F^{\mu\nu}$. The first two can be subtracted off with the appropriate counter-terms, and the third is a total derivative. Hence, the expressions \eqref{eqn:effectivelagrangianeven1} and \eqref{eqn:effectivelagrangianodd1} hold without loss of generality. 

To conclude, we have derived the even and odd parts of the full non-perturbative effective potential for QED with general Yukawa couplings, valid to all orders in the background scalar, pseudoscalar, and electromagnetic fields. More precisely, it is the zeroth-order result in the derivative expansion of the full effective Lagrangian. The even portion is simply the Euler-Heisenberg effective action but with the formal  replacement $m^2 \to (m+\kappa\Phi)^2+(\lambda\Pi)^2,$ which is expected from the Dirac structure of the Yukawa couplings. The even part only depends on $\Pi^2$, and hence generates graphs with even numbers of pseudoscalar vertices. The odd portion is (somewhat surprisingly) proportional to $F\tilde F$ with no higher-order corrections appearing.   

In the limit $\Phi ,\Pi \to 0$, the odd portion vanishes and we exactly recover the Euler-Heisenberg Lagrangian. Similarly, if we let $A_\mu \to 0$, the odd portion vanishes, but the even portion becomes 
\begin{equation}
\left. \mathscr{L}_{\text{eff}}\right|_{A_\mu=0} = -\frac{1}{8\pi^2}\int_0^\infty \frac{ds}{s^3} e^{-s((m+\kappa\Phi)^2+(\lambda\Pi)^2)}.
\end{equation}
The bad behavior of the integral as $s \to 0$ reflects an ultraviolet divergence. As we will see, the divergence can be handled with appropriate counter-terms.


\section{Weak Field Limit and Renormalization}

In this section we consider the perturbative expansion of \eqref{eqn:effectivelagrangian1} in powers of $e$. We begin with the even part, given by \eqref{eqn:effectivelagrangianeven1}.
 Grouping by powers of $s$, and using \eqref{eqn:lorentzinvariants2}, the weak-field expansion becomes
\begin{subequations}
\begin{eqnarray}
\mathscr L_{\textrm{eff}}^{\text{even}} &=&-\frac{1}{8 \pi^2} \int_0^\infty \frac{ds}{s^3} e^{-s(\tilde m^2)} - \frac{e^2}{48\pi^2} F_{\mu\nu}F^{\mu\nu} \int_0^\infty \frac{ds}{s} e^{-s(\tilde m^2)} \label{eqn:weakfield1} \\
&&+ \frac{e^4}{1440\pi^2} \left( (F_{\mu\nu}F^{\mu\nu})^2 + \frac{7}{4}(F_{\mu\nu}\tilde F^{\mu\nu})^2 \right)\int_0^\infty ds s\  e^{-s(\tilde m^2)} + \mathcal O(e^6), 
\end{eqnarray}
\end{subequations}
which is manifestly gauge-invariant. For ease of notation we have introduced
\begin{equation}
\tilde m \equiv \sqrt{(m+\kappa\Phi)^2+(\lambda\Pi)^2}.
\end{equation}
The first two integrals \eqref{eqn:weakfield1} are badly behaved as $s\to0$, and require regularization. This can be achieved by cutting off the lower bound of the integral at a small positive number $\epsilon > 0$ (making use of the incomplete Gamma function), or by analytic continuation of the complete Gamma function. We take the latter approach. Despite its resemblance to dimensional regularization, this regularization procedure does not analytically continue the number of spacetime dimensions, but rather the power of $s$ in the denominator of the integrand, which is not physical. Making use of the integral representation of the Gamma function, 
\begin{equation}
\int_0^\infty \frac{ds}{s^{1-z}} e^{-s\alpha} = \frac{\Gamma(z)}{\alpha^z},
\end{equation}
we can compute the integrals above. The result is
\begin{eqnarray}
\mathscr L_{\textrm{eff}}^{\text{even}} &=& -\frac{\tilde m^4}{16 \pi^2}\left( \frac{1}{\epsilon} - \gamma + \frac{3}{2} - \ln \tilde m^2 \right) - \frac{e^2}{48\pi^2} \left(\frac{1}{\epsilon}-\gamma-\ln \tilde m^2\right)F_{\mu\nu}F^{\mu\nu}  \nonumber\\
&&+ \frac{e^4}{1440\pi^2\tilde m^4} \left( (F_{\mu\nu}F^{\mu\nu})^2 + \frac{7}{4}(F_{\mu\nu}\tilde F^{\mu\nu})^2 \right) + \mathcal O(e^6),
\end{eqnarray}
where $\gamma$ is the Euler-Mascheroni constant and the limit $\epsilon\to0$ is assumed. Now we proceed in the $\overline{\text{MS}}$ scheme, adding two counter terms
\begin{subequations}
\begin{eqnarray}
\mathscr L_{\text{c.t.}}^{(0)} &=& \frac{\tilde m^4}{16\pi^2} \left(\frac{1}{\epsilon}-\gamma\right)M^{-2\epsilon}, \\
\mathscr L_{\text{c.t.}}^{(2)} &=& \frac{e^2}{48\pi^2} \left(\frac{1}{\epsilon}-\gamma\right)M^{-2\epsilon} F_{\mu\nu}F^{\mu\nu},
\end{eqnarray}
\end{subequations}
where we have introduced the renormalization scale $M$ to ensure the counter terms have the correct dimension. The first counter term corresponds to the renormalization of the cosmological constant and the scalar potential $V[\Phi,\Pi]$; specifically, the $\Phi, \Phi^2, \Phi^3,\Phi^4,\Pi^2,\Pi^4,\Pi^2\Phi,$ and $\Pi^2\Phi^2$ couplings. The second counter term corresponds to the renormalization of the free photon term and is related to the vacuum polarization. The $\epsilon \to 0$ limit of the sum $\mathscr L_{\text{eff}} + \mathscr L_{\text{c.t.}}^{(0)} + \mathscr L_{\text{c.t.}}^{(2)}$ is finite, 
\begin{eqnarray}
\mathscr L_{\text{eff}} + \mathscr L_{\text{c.t.}} &=& \frac{\tilde m^4}{16\pi^2} \left( \ln \frac{\tilde m^2}{M^2} - \frac{3}{2} \right) + \frac{e^2}{48\pi^2} \ln \frac{\tilde m^2}{M^2} F_{\mu\nu}F^{\mu\nu} \nonumber\\ 
&&+  \frac{e^4}{1440\pi^2\tilde m^4} \left( (F_{\mu\nu}F^{\mu\nu})^2 + \frac{7}{4}(F_{\mu\nu}\tilde F^{\mu\nu})^2 \right) + \mathcal O(e^6).
\end{eqnarray}
We introduce $\bar{M}^2 = M^2e^{3/2}$ (here $e$ is Euler's constant not the electric charge) and the fine structure constant $\alpha_0 = e^2/4\pi$. Including the free photon term from the bare Lagrangian, the even portion of the full renormalized effective Lagrangian $\mathscr{L}^{\text{even}}_{\text{1-loop}} = \mathscr L_{\text{bare}} + \mathscr L_{\text{eff}}^{\text{even}}+ \mathscr L_{\text{c.t.}}$ to first order in the nonlinear effects is
\begin{eqnarray}\label{eqn:EHweakfield}
\mathscr{L}^{\text{even}}_{\text{1-loop}} = \frac{\tilde m^4}{16\pi^2} \ln \frac{\tilde m^2}{\bar{M}^2}&&-\frac{1}{4}F_{\mu\nu}F^{\mu\nu}\left(1- \frac{\alpha_0}{3\pi}  \ln \frac{\tilde m^2}{M^2}\right) \nonumber  \\
&&+  \frac{\alpha_0^2}{90\tilde m^4} \left( (F_{\mu\nu}F^{\mu\nu})^2 + \frac{7}{4}(F_{\mu\nu}\tilde F^{\mu\nu})^2 \right).
\end{eqnarray}
One can, at this stage, consider the limit in which the electromagnetic fields vanish. We then find
\begin{equation}
\left.\mathscr{L}_{\text{eff}}\right|_{A_\mu=0} = \frac{\left((m+\kappa\Phi)^2+(\lambda\Pi)^2\right)^2}{16\pi^2} \ln \frac{(m+\kappa\Phi)^2+(\lambda\Pi)^2}{\bar{M}^2},
\end{equation}
which is reminiscent of the Coleman-Weinberg potential for scalar electrodynamics \cite{PhysRevD.7.1888}.

The odd part of the effective potential is finite. Of course, in order for the underlying theory to be one-loop renormalizable, this must be the case. Expanding to first order in the pseudoscalar coupling, and letting $\Phi = 0$ for simplicity, \eqref{eqn:effectivelagrangianodd1} becomes 
\begin{equation}
\mathscr{L}_{\text{eff}}^\text{odd} = \frac{\alpha_0}{4\pi}F\tilde F\arctan\left(\frac{\lambda\Pi}{m+\kappa\Phi}\right) \approx \frac{\alpha_0}{4\pi}\frac{\lambda}{m}\Pi F\tilde F. 
\end{equation}
This shift-symmetric axion-type interaction between the pseudoscalar $\Pi$ and $F\tilde F$ has been well-studied in the context of cosmological inflation. For instance, such a coupling arises in models of natural inflation \cite{PhysRevLett.65.3233,Barnaby:2011vw,Anber:2009ua,Papageorgiou:2017yup,Bartolo:2017szm,Sorbo:2011rz}. While these effective interactions are usually assumed to arise from some fundamental theory (often a string theory), here we have shown how such a coupling can arise from a simple renormalizable extension of QED.


\section{Chiral Anomaly}

Finally, as an application of our result, we show how the chiral anomaly in QED can be readily connected to the pseudoscalar coupling in $\mathscr L_{\text{YUK}}$ and the odd part of the effective action \eqref{eqn:effectivelagrangianodd1}. While there is no axial symmetry present in the full model with Yukawa couplings \eqref{eqn:lagrangian}, the classical QED action exhibits an axial symmetry in the limit that the electron mass vanishes. Classically, the associated axial current $J^{\mu 5} = \bar\psi\gamma^\mu\gamma^5\psi$ satisfies the following relation, 
\begin{equation}
\partial_\mu J^{\mu 5} = 2im\bar\psi\gamma^5\psi,
\end{equation}
which can be obtained by using the equations of motion for the Dirac field,
\begin{subequations}
\label{equations}
\begin{eqnarray}
i\slashed\partial\psi &= m\psi-e\slashed A\psi, \\
-i\slashed\partial\bar\psi &= m\bar\psi-e\slashed A\bar\psi.
\end{eqnarray}
\end{subequations}
Hence, in the limit that $m\to 0$, the axial current is classically conserved. It is well-known, however, that the divergence of the axial current is anomalous, 
\begin{equation}
\partial_\mu J^{\mu 5} = 2im\bar\psi\gamma^5\psi + \frac{\alpha_0}{2\pi} F_{\mu\nu}\tilde F^{\mu\nu}.
\end{equation}
The second term, which arises when the quantum effects are fully taken into account, is the Adler-Bell-Jackiw anomaly \cite{adler,belljackiw,PhysRevD.21.2848,Dittrich:2000az,Dittrich:2000zu}. Somewhat surprisingly, in the case of a constant field strength the contribution from the explicit symmetry breaking term proportional to $m$ cancels with the anomaly:
\begin{equation}
\langle A_\mu |\partial_\mu J^{\mu 5}|A_\mu \rangle = 2im\langle A_\mu |\bar\psi\gamma^5\psi|A_\mu \rangle + \frac{\alpha_0}{2\pi} F_{\mu\nu}\tilde F^{\mu\nu}=0,
\end{equation}
implying that
\begin{equation}
\langle A_\mu |\bar\psi\gamma^5\psi|A_\mu \rangle = i\frac{\alpha_0}{4\pi m}F_{\mu\nu}\tilde F^{\mu\nu}.
\end{equation}
This fact can be shown via Schwinger's equivalence theorem \cite{Schwinger, Dittrich:2000zu}, and reveals that the axial current (which is classically non-conserved due to the mass of the fermion) is actually conserved in the quantum theory. 

In the effective theory defined by \eqref{eqn:effectivelagrangian1}, the expectation value of the chiral density in the classical electromagnetic background is given by the pseudoscalar current, which can be be read off directly from \eqref{eqn:oddcurrent}. We find
\begin{equation}
\langle A_\mu |\bar\psi\gamma^5\psi|A_\mu \rangle =- \left.\frac{1}{i}\frac{1}{\lambda} \frac{\delta \Gamma_{\text{1-loop}}}{\delta \Pi}\right|_{\Phi,\Pi=0} = -\left. J_\Pi\right|_{\Phi,\Pi=0} = i\frac{\alpha_0}{4\pi m}F_{\mu\nu}\tilde F^{\mu\nu}, 
\end{equation}
and hence, in the constant electromagnetic background, $\partial_\mu J^{\mu 5} = 0$ when both the explicit symmetry breaking term and the anomaly are taken into account.


\section{Conclusion}

We have presented a derivation of the effective action for QED with Yukawa couplings in the one-loop and constant background field approximations. Rather than computing the fermion determinant directly, we have derived the contribution of the effective action to the pseudoscalar current, which is then used to reconstruct the full one-loop effective Lagrangian. In doing so, we avoided subtleties regarding the Hermiticity of the generalized Dirac operator in the underlying theory, which involves $\gamma^5$. 

Using the Fock-Schwinger proper-time formalism and zeta-function regularization, we obtained an effective action with an even and odd part. The even portion reduces to the well-known Euler-Heisenberg effective action in the limit that the scalar and pseudoscalar fields vanish, while the odd portion vanishes identically when either the pseudoscalar or vector gauge fields are taken to zero. Our result is non-perturbative in the background fields, and unlike the world-line path integral representation given in \cite{DHoker:1995aat}, it clearly generalizes the closed-form Euler-Heisenberg effective action. While our result is derived assuming the scalar and pseudoscalar fields are distinct so that the original Lagrangian is parity invariant, it is also valid when $\Phi$ and $\Pi$ are identified as the same field. In this case, the original Lagrangian explicitly breaks parity, and parity-breaking radiative corrections are induced. 

The odd part of the effective action is proportional to $F\tilde F$, with no higher-order odd powers appearing. This suggests that couplings involving higher powers of $F\tilde F$ and $\Pi$ appear only when derivative corrections are included. While the even portion is invariant under global chiral transformations, the odd portion is related to the phase angle between the scalar and pseudoscalar fields. For large values of the pseudoscalar field, or in the limit that the fermion mass vanishes, the odd portion of the effective action approaches a total derivative. The first non-trivial term in the weak-field expansion is an axion-like interaction, which is of phenomenological interest in the context of cosmology. We also find that the chiral anomaly in QED cancels against the explicit symmetry breaking term proportional to the fermion mass, which can be easily obtained from the odd part of the pseudoscalar current. The interplay between the anomaly and the explicit symmetry breaking term at finite temperature, which controls the thermal $\pi^0 \to \gamma\gamma$ decay, can be connected to the finite temperature effective action of the QED-Yukawa theory, which will be treated in a future work.

\begin{acknowledgments}
The authors wish to thank T.E. Clark for useful discussions. The work of T.N.J. was supported by the Beltmann Physical Sciences Summer Research Fund at Macalester College.  
\end{acknowledgments}

\bibliographystyle{utphys}
\bibliography{EffectiveActionARXIV}

\end{document}